\documentclass[runningheads]{llncs}
\usepackage{amsmath}
\usepackage[T1]{fontenc}
\usepackage{graphicx}
\usepackage{listings}
\usepackage{amsmath}
\usepackage{amssymb}
\usepackage{graphicx}
\usepackage{bm}
\usepackage{color}
\usepackage{authblk}
\usepackage{booktabs}
\usepackage{multirow}
\usepackage{float}
\usepackage{tikz}
\usepackage{ctable}
\usepackage{fontawesome}
\usepackage[colorlinks, citecolor=green]{hyperref}

\newcommand\tstrut{\rule{0pt}{2.4ex}}

\usepackage{color, colortbl}
\definecolor{Gray}{gray}{0.9}
\usetikzlibrary{arrows.meta,arrows}
\usepackage{diagbox}
\usepackage{breakcites}
\usepackage{orcidlink}
\usepackage{setspace}
\usepackage{amsmath}
\begin{document}
\title{Classification, Regression and Segmentation directly from k-Space in Cardiac MRI}
\titlerunning{K-Space Comparative Analysis}
%
\author{Ruochen Li$^*$\textsuperscript{1(\faEnvelopeO{})} \and
        Jiazhen Pan$^*$\inst{1} \and
        Youxiang Zhu\inst{2} \and 
        Juncheng Ni\inst{1} \and 
        Daniel Rueckert\inst{1,3}}


\authorrunning{R. Li et al.}

\institute{
    Technical University of Munich, Germany \\
    \email{ruochen.li@tum.de}
    \and
    University of Massachusetts Boston, Boston, MA, USA
    \and
    BioMedIA, Imperial College London, United Kingdom
}

\maketitle              
\def\thefootnote{*}\footnotetext{Equal contribution}
\begin{abstract}
Cardiac Magnetic Resonance Imaging (CMR) is the gold standard for diagnosing cardiovascular diseases. Clinical diagnoses predominantly rely on magnitude-only Digital Imaging and Communications in Medicine (DICOM) images, omitting crucial phase information that might provide additional diagnostic benefits. In contrast, k-space is complex-valued and encompasses both magnitude and phase information, while humans cannot directly perceive. In this work, we propose KMAE, a Transformer-based model specifically designed to process k-space data directly, eliminating conventional intermediary conversion steps to the image domain. KMAE can handle critical cardiac disease classification, relevant phenotype regression, and cardiac morphology segmentation tasks. We utilize this model to investigate the potential of k-space-based diagnosis in cardiac MRI. Notably, this model achieves competitive classification and regression performance compared to image-domain methods e.g. Masked Autoencoders (MAEs) and delivers satisfactory segmentation performance with a myocardium dice score of 0.884. Last but not least, our model exhibits robust performance with consistent results even when the k-space is 8× undersampled. We encourage the MR community to explore the untapped potential of k-space and pursue end-to-end, automated diagnosis with reduced human intervention. Codes are available at \url{https://github.com/ruochenli99/KMAE_cardiac}.

\end{abstract}

%
%
%
\section{Introduction}

Cardiac Magnetic Resonance Imaging (CMR) serves as the gold standard for diagnosing and treating cardiovascular diseases, offering a comprehensive view of the heart's morphology and function. This non-invasive method enables detailed assessments of myocardial viability, ventricular function, and vascular anatomy. While Digital Imaging and Communications in Medicine (DICOM) protocol images are the prevalent format for storage and visualization, they consist solely of magnitude data derived from the real and imaginary components of the original complex data. Crucially, the phase information omitted in DICOM images holds potential value for tasks such as image reconstruction, segmentation, and the evaluation of flow dynamics and tissue movement~\cite{sriram2020end,hajivalizadeh2021comparison,rempe2024kstrip}. Meanwhile, accelerated MR scans are preferred in clinics to reduce scan time and enhance patient comfort, leading to undersampled k-space(frequency domain representation of MR signal), which results in corrupted/blurred CMR images and deteriorates the follow-up downstream tasks~\cite{hammernik2021motion,pan2022learning,gong2022robust,lyu2024state,pan2024reconstruction}.

Recently, methods that perform downstream tasks, e.g., motion estimation~\cite{kustner2021lapnet,kustner2022self} and segmentation~\cite{schlemper2018cardiac,rempe2024kstrip,zhang2024direct} directly from k-space data gained attention. K-space, being complex-valued, encapsulates phase information and remains an intact and reliable data source with no corruptions, despite some acquisition lines that could be missing in undersampling. However, humans may struggle to perceive k-space data since they are not visually understandable to humans. Conversely, deep learning models excel in processing these data, as their computational frameworks readily handle complex values. Given the complexity and rich content of k-space data, selecting appropriate methods to effectively process and utilize this data is crucial for optimizing the diagnostic capabilities of cardiac MRI and for a comprehensive assessment of cardiovascular health.

Transformers are highly proficient in capturing long-range dependencies~\cite{li2022modeling} and handling complex data structures, making them well-suited for modeling the temporal dynamics and global information present in k-space data~\cite{oh2021kspace}. Pan et al. proposed the Transformer-based K-GIN model~\cite{pan2023global}, showing outstanding performance in MRI reconstruction solely using k-space data, highlighting the strong capabilities of its encoders in feature extraction and representation learning. We argue that this learned representation is not limited to the reconstruction tasks, but can be leveraged to more diverse tasks such as classification and segmentation. In this work, we propose KMAE, a versatile model that takes (undersampled) k-space data as inputs and can handle various downstream tasks, including disease classification, relevant phenotype regression, and cardiac segmentation. It leverages the pre-trained K-GIN encoders to attain rich representation and applies different decoders to carry out diverse downstream tasks. This adaptation facilitates efficient and accurate diagnostics and analyses based on k-space data. The contributions of this study can be summarised as follows:

\begin{enumerate}
    \item We propose KMAE, a Transformer-based method for processing cardiac MR k-space data. KMAE can perform multiple downstream tasks, including disease classification, phenotype regression, and cardiac segmentation. To the best of our knowledge, we are the first to conduct disease classification directly from k-space data.
    \item Unlike Convolutional Neural Networks (CNNs), which use local convolutional windows, we demonstrate that Transformers, which capture long-range dependencies, are more effective and robust for k-space data.
    \item KMAE achieves competitive classification and regression performance compared to image-domain methods such as Masked Autoencoders (MAEs). It also provides satisfactory segmentation results with a myocardium dice score of 0.884, matching the quality of image-domain segmentation. Our model exhibits robust performance, maintaining consistent results even with 8× undersampled k-space data.
\end{enumerate}

\section{Related Work}

\noindent \textit{\textbf{K-space Interpolation:}} Previous methods typically leverage auto-calibration signals (ACS) in the k-space center to carry out k-space interpolation~\cite{GRAPPA,lustig2010spirit}. RAKI~\cite{akcakaya2019raki,kim2019loraki} improved the ACS-based methods by implementing CNNs. However, these approaches did not fully exploit the global dependencies present in k-space. Recently, a Transformer-based k-space interpolation method considering k-space global dependencies for dynamic CMR reconstruction was introduced by Pan et al.~\cite{pan2023global}, achieving superior performance compared to baselines.

\noindent \textit{\textbf{Downstream Tasks directly from K-space:}} Schlemper et al.~\cite{schlemper2018cardiac} proposed CNN-based models with an end-to-end synthesis network and a latent feature interpolation network, predicting cardiac segmentation maps directly from undersampled dynamic MRI data. Kuestner et al. introduced LAP-Net~\cite{kustner2021lapnet}, which can estimate the cardiac motion from the k-space of Cardiac MR. Moritz Rempe et al. proposed k-strip model~\cite{rempe2024kstrip}, a complex-valued CNN-based algorithm for skull stripping in MRI, skipping operations in the image domain. Nevertheless, these methods are built upon CNNs and may not be able to fully exploit the global dependencies in k-space. Concurrently, Zhang et al.~\cite{zhang2024direct} proposed to use Transformers to directly derive segmentation from undersampled k-space data.

\noindent \textit{\textbf{Masked Image Modelling:}} Vision Transformers(ViTs)~\cite{dosovitskiy2020image} adapted Transformers from natural language processing to computer vision. Unlike CNNs that rely on local convolutions, the global self-attention mechanism of ViTs allows for the modeling of long-range dependencies within images~\cite{islam2023comprehensive}; Masked Autoencoders (MAEs)~\cite{he2022masked}, was introduced based on ViTs to extract the representation using masked image modeling in a self-supervised manner. Its versatility is further demonstrated in cardiac MR imaging analysis~\cite{zhang2024whole}. Recently, K-GIN~\cite{pan2023global} was introduced based on MAEs to learn k-space representation and conduct cardiac MR reconstruction, presenting robust and superior performance. We argue that its learned representation is not limited to the reconstruction tasks, but also the other tasks such as classifications and segmentation.

\begin{figure}[t]
\centering 
\includegraphics[width=1.0\textwidth]{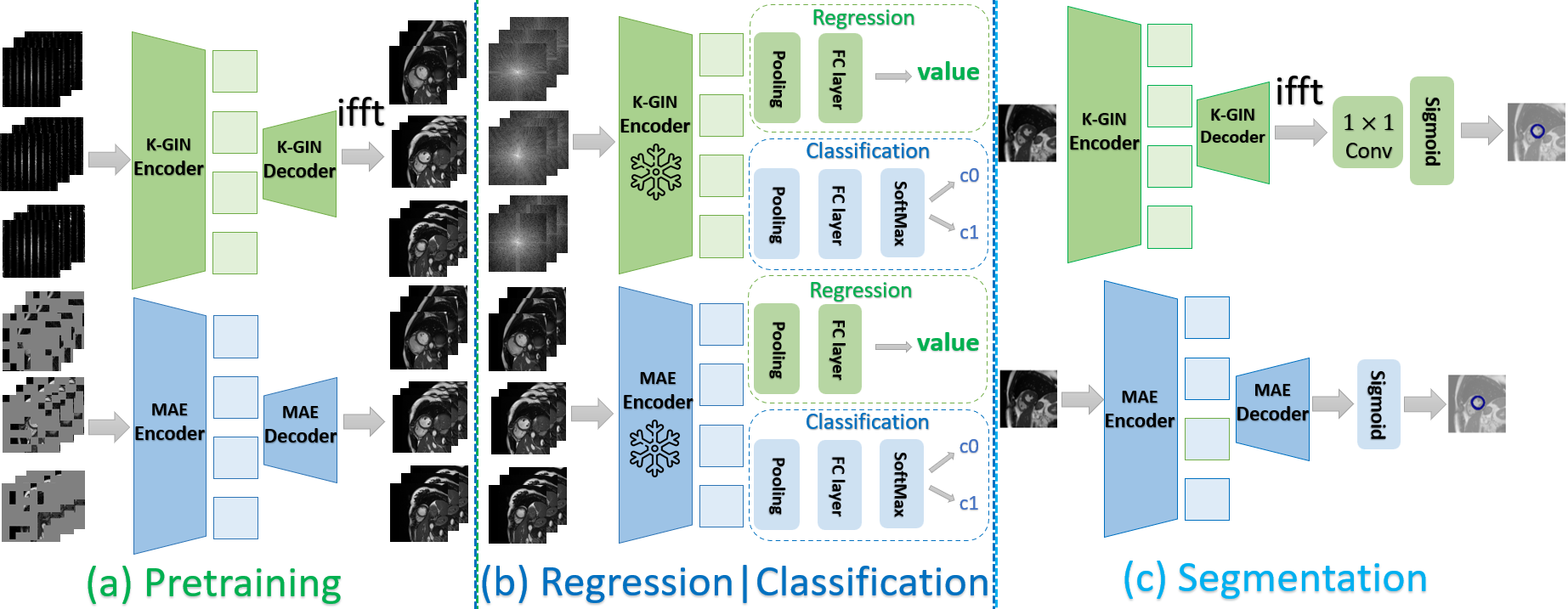} 
\caption{An overview of KMAE and MAEs with downstream tasks. The upper section depicts KMAE, a modification of the K-GIN model. The lower section illustrates the modification of MAEs. (a) The pre-training of KMAE and MAEs for MRI reconstruction. KMAE processes under-sampled k-space data, while MAEs handle in the image domain with masked-out patches. (b) In downstream task fine-tuning, we freeze encoders of KMAE and MAEs while their decoders are modified for regression and classification. (c) We adapt decoders of KMAE and MAEs for segmentation tasks, with the upper section highlighting our newly proposed k-space segmentation method.} 
\label{fig:Structure} 
\end{figure}

\section{Methods}

\subsection{Pre-Training}
\noindent \textit{\textbf{Models:}} \textbf{K-GIN}~\cite{pan2023global}, designed for MRI reconstruction, processes undersampled k-space data (e.g., Cartesian undersampling) and performs k-space interpolation to predict fully sampled k-space data, and converts it to MRI via an inverse fast Fourier transform. \textbf{MAEs}~\cite{he2022masked} features an asymmetric design. The model inputs images with most patches masked, exposing only a few. The encoder processes these visible patches and passes them to a smaller decoder, whose primary task is reconstructing the original image pixels, resulting in high-quality MRI images.

\noindent \textit{\textbf{Pre-Training Process:}} We utilize k-space interpolation / image reconstruction tasks to pre-train KMAE / MAEs, as illustrated in Figure~\ref{fig:Structure}(a). During the pre-training phase, our KMAE model rigorously adheres to the foundational principles and procedures established by the K-GIN architecture.

\noindent \textit{\textbf{Evaluation Metrics Meaning:}} After pre-training, both KMAE and MAEs achieve high Peak Signal-to-Noise Ratio (PSNR) values, showcasing the excellent quality of their reconstructed images. This performance illustrates their encoders' effectiveness in extracting meaningful representations from raw data, facilitating subsequent tasks. Moreover, this pre-training approach significantly reduces training times, enabling faster adaptation to various downstream tasks.

\subsection{Regression and Classification}
After pre-training, we freeze the model's encoder and discard the reconstruction decoder. The trained encoder is then used to extract valuable feature representations for downstream tasks such as regression and classification, as shown in Figure~\ref{fig:Structure}(b). Both KMAE and MAEs adopt a consistent architectural framework for downstream tasks.

For \textbf{regression} tasks, we employ a pooling layer to transform the extracted features into a feature vector. This vector is then fed into a fully connected (FC) layer to predict the regression value for each subject. Regarding \textbf{classification} tasks, we incorporate a final layer equipped with a SoftMax function, which computes the probability of each class to categorize the subjects.

\subsection{Segmentation}

In regression and classification tasks, outputs are numerical values and probabilities, so the decoder used for image reconstruction in pre-training is removed as it is unnecessary. However, image segmentation tasks remain closely linked to reconstruction tasks as they both require pixel-level prediction. Therefore, we adopted the same reconstruction decoder to accomplish CMR segmentation.

Regarding \textbf{MAEs}, it processes in the image domain and we can adapt it to segmentation tasks by simply replacing the final layer with a Sigmoid function. On the other hand, \textbf{KMAE} handles in the frequency domain, therefore we first Fourier transfer the reconstructed k-space to MR images. These images are then processed through a 1x1 convolution layer, followed by a Sigmoid function to effectively segment the myocardium. Both structures are described in Figure~\ref{fig:Structure}(c).

\section{Data and Experiments }
\subsection{Dataset}
\noindent \textit{\textbf{Dataset.}} We used short-axis cardiac MR images provided by UKBioBank~\cite{petersen2015uk} and corresponding clinical information, which provide a cross-sectional view of the left and right ventricles of the heart. We applied center-cropped CMR images with a matrix size of 128×128 across 25 temporal cardiac phases (we used every two temporal frames). Since the original CMR from UKBioBank are magnitude-only images, we created synthetic k-space data for each 2D+time scan by applying additional Gaussian B0 variations in real-time to remove the conjugate symmetry of k-space \cite{schlemper2018cardiac}, thus simulating fully sampled single-coil acquisitions. We stacked 11 slices along the long axis. Additionally, we applied VISTA Cartesian undersampling masks \cite{Vista} to generate the accelerated k-space and the corresponding MRI.

\noindent \textit{\textbf{Filter Data and Label Strategy.}} The UK Biobank CMR dataset initially comprises 47,097 subjects. We identified three distinct subsets for our study: \textbf{Healthy Subgroup}, consisting of 2,660 individuals without risk factors such as obesity, myocardial infarction, acute myocardial infarction, insulin-dependent diabetes mellitus, or physician-diagnosed vascular or heart conditions. This subgroup only includes individuals rated as "Excellent" or "Good" in overall health who also reported never having smoked tobacco~\cite{shah2023environmental}. \textbf{Cardiopathy Subgroup}~\cite{clough2019global} comprises 1,340 subjects with diagnosed heart conditions, including heart attacks, myocardial infarction, and angina. \textbf{Left Ventricular Dysfunction Subgroup}~\cite{elghazaly2023characterizing} includes 937 subjects with a Left Ventricular Ejection Fraction (LVEF) below 50\%. For \textbf{regression}, we selected 2,000 subjects from the Healthy Subgroup to calculate cardiac age based on birth year and scan date~\cite{inacio2023cardiac}. We used LVEF and LVEDV(Left Ventricular End-Diastolic Volume) labels from 1,000 healthy subjects sourced from~\cite{bai2020populationbased}. For \textbf{classification}, we compared 937 subjects from the Left Ventricular Dysfunction Subgroup to an equal number from the Healthy Subgroup. Similarly, 1,340 subjects from the Cardiopathy Subgroup were matched with an equivalent number from the Healthy Subgroup.

\subsection{Implementing details}

\noindent \textit{\textbf{Pre-training.}} We trained MAEs and K-GIN on data from the Healthy Subgroup, which consisted of MRI datasets with 5 slices and 25 temporal frames. Both models were tasked with image reconstruction, achieving PSNR of 38.846 for MAEs and 38.755 for K-GIN. The MAEs used a patch size of 2, while all other hyperparameters remained consistent with the original MAE specifications. Similarly, K-GIN adhered to its original configurations. Details of the implementation are disclosed in our code repository.

\noindent \textit{\textbf{Training Strategy.}} We employed an NVIDIA A40 GPU to train our framework, configuring the setup with a single batch and a learning rate scheduler, peaking at 0.0001. Our Transformer architecture utilized 8 layers, 8 heads, and an embedding dimension of 512, while the ResNet model was trained without pre-trained weights from cardiac MRI data. For \textbf{classification} and \textbf{regression} tasks, we processed 5 MRI slices per subject, each containing 25 frames, and averaged the results from each slice to compute final regression scores or classification probabilities. The KMAE and MAEs' encoder were frozen, with only training on subsequent layers, as shown in Figure~\ref{fig:Structure}(b). Moreover, we performed a comprehensive performance comparison by training the full KMAE pipeline without freezing any components. For \textbf{segmentation} tasks, we used a single MRI slice with 25 frames per subject to accurately segment myocardial regions with no encoder freezing, as illustrated in Figure~\ref{fig:Structure}(c).

\noindent \textit{\textbf{ResNet Baseline.}} Our k-space data includes 2D spatial and temporal dimensions (2D+t). So, we adapted ResNet50 by modifying the channel dimensions of its 2D convolutional layers to match the number of cardiac SAX slices.

\begin{table}[t]

\centering
\begin{tabular}{p{2.7cm} p{1.5cm} p{1.5cm} p{2cm} p{2cm} p{2cm}}
\toprule
& \multicolumn{3}{c}{Regression} & \multicolumn{2}{c}{Classification} \\ 
& Age $\downarrow$ & LVEF $\downarrow$ & LVEDV $\downarrow$ & LV Dys $\uparrow$ & Cardiopathy$\uparrow$ \\
\rowcolor{Gray}
\hline
ResNet & 6.031 & 5.887 & 27.188 & 63.31\% & 72.54\% \tstrut\\
ResNet(R=4) & 6.559 & 5.443 & 22.012 & 65.09\% & 72.95\% \tstrut\\
\rowcolor{Gray}
KMAE & 5.840 & 4.547 & 23.917 & 68.64\% & 75.41\% \tstrut\\
KMAE(R=4) & 5.690 & 4.568 & 22.591 & 69.82\% & 75.00\% \tstrut\\
\rowcolor{Gray}
KMAE(R=4,\dag) & \textbf{4.439} & \textbf{4.128} & \textbf{16.452} & \textbf{76.33\%} & \textbf{77.46\%} \tstrut\\
\bottomrule
\end{tabular}
\caption{Comparison of different models using two types of inputs: using original k-space data (first and third rows) and using undersampled k-space data (second and fourth rows). The evaluation metrics include Mean Absolute Error for regression and accuracy for classification. R=4 denotes the acceleration rate for undersampling k-space. '\dag' means the encoder is not frozen. The best results are marked in bold.} 
\label{table:comp1}
\end{table}

\begin{table}[t]

\centering
\begin{tabular}{p{2.5cm} p{1.5cm} p{1.5cm} p{2cm} p{2cm} p{2cm}}
\toprule
& \multicolumn{3}{c}{Regression} & \multicolumn{2}{c}{Classification} \\
& Age $\downarrow$ & LVEF $\downarrow$ & LVEDV $\downarrow$ & LV Dys $\uparrow$ & Cardiopathy$\uparrow$\\
\rowcolor{Gray}
\hline
MAEs & \textbf{5.553} & \textbf{4.511} & 23.545 & \textbf{78.36}\% & \textbf{76.45\%} \tstrut\\
KMAE & 5.840 & 4.547 & 23.917 & 68.64\% & 75.41\% \tstrut\\
\rowcolor{Gray}
KMAE(R=4) & 5.690 & 4.568 & \textbf{22.591} & 69.82\% & 75.00\% \tstrut\\
KMAE(R=8) & 5.669 & 4.610 & 22.694 & 69.23\% & 74.59\% \tstrut\\
\bottomrule
\end{tabular}
\caption{Comparison of one model using three types of inputs: using original MRI image (first row), original k-space data (second row), and using undersampled k-space data (third and fourth rows). The metrics evaluated include Mean Absolute Error for regression and accuracy for classification. The best results are marked in bold.}
\label{table:comp2}
\end{table}


\noindent \textit{\textbf{Metrics.}} In accelerated CMR, where CMR imaging employs acceleration techniques, higher acceleration factors (R=4 or R=8) lead to increased undersampling of k-space data. For \textbf{regression} tasks, we used Huber loss to train and evaluated performance by Mean Absolute Error (MAE). Lower MAE values indicate better regression performance. For \textbf{classification} tasks, performance was assessed using cross-entropy loss and accuracy, with higher accuracy indicating better classification results. For \textbf{segmentation} task, we employed binary cross-entropy loss during training and gauged effectiveness with the Dice score, with higher scores indicating enhanced segmentation performance.

\section{Results and Discussion}

In table \ref{table:comp1}, KMAE generally exhibits lower MAE values for regression tasks, indicating more accurate predictions for variables such as age, LVEF, and LVEDV. Even with undersampling k-space data, KMAE tends to outperform ResNet. KMAE consistently achieves higher accuracy for classification tasks than ResNet, regardless of undersampling or freezing layers. 

Table \ref {table:comp2} demonstrates that the Transformer model performs comparably well with undersampled k-space data as input, even when compared to original MRI images. This adaptability is evident in the first two rows of the table. Even when undersampled k-space data is used as input (KMAE at R=4 and KMAE at R=8), the model still achieves competitive performance, with only slight variations compared to the implementation on the full sampled k-space data.

Table \ref{table:comp3} shows that MAEs achieves the highest Dice coefficient. Meanwhile, KMAE and its undersampled version also exhibit reasonably high Dice coefficients, demonstrating that they are capable of producing accurate segmentation results, albeit slightly lower than the MAEs model.

\begin{table}[t]
\centering
\setlength{\tabcolsep}{3.5mm}{}
\begin{tabular}{lcccc}
\toprule
& MAEs & KMAE & KMAE(R=4) & KMAE(R=8) \\\rowcolor{Gray}
\hline
DICE & 0.941 & 0.884 & 0.873 & 0.870 \\
\bottomrule
\end{tabular}
\caption{Comparison of one model using three types of inputs for segmentation task: using original MRI image(first column), original k-space data (second column), and using undersampled k-space data (third and fourth columns). }
\label{table:comp3}
\end{table}

\begin{figure}[t]
\centering 
\includegraphics[width=1.0\textwidth]{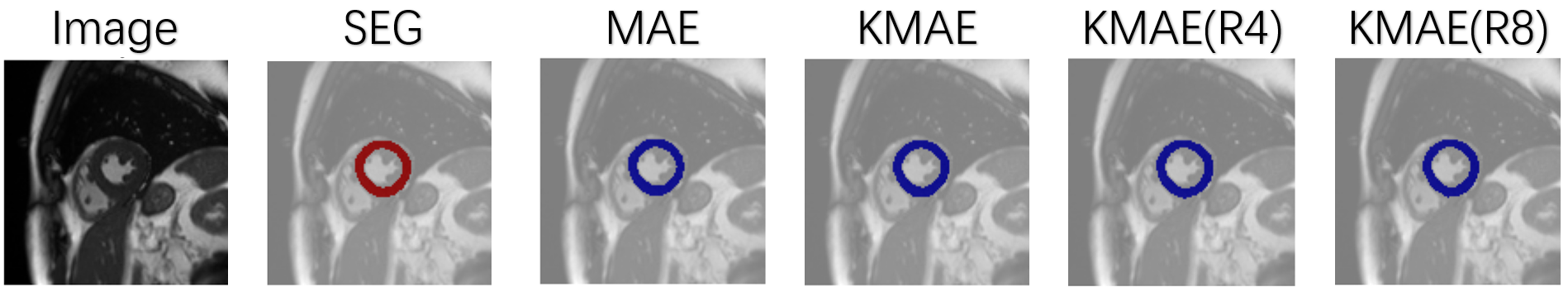} 
\caption{Comparison of segmentation methods for delineating myocardial regions  }
\label{fig:seg_result} 
\end{figure}

Figure \ref{fig:seg_result} shows that MAEs, utilizing CMR images as the input, delivers optimal segmentation performance by precisely delineating the myocardium within the heart. Furthermore, KMAE employing undersampled k-space inputs also exhibits impressive segmentation capabilities.

Our experiments show that the Transformer architecture is highly effective for processing k-space data, particularly when undersampled. The self-attention mechanism in Transformers efficiently handles global information in k-space, in which each k-space measurement point contributes to forming every pixel in the image domain. This capability enables the model to assimilate information from diverse positions, effectively capturing global correlations. Specifically in cardiac applications, Transformers treats k-space frames as time sequences, integrating temporal information by segmenting frames into patches, thus capturing cardiac dynamics over time. Notably, predictions derived from undersampled k-space outperform those from fully-sampled k-space, as the latter often includes irrelevant details to classification / segmentation. At the same time, the former focuses more on low-frequency components, which is the most critical information contributing to the downstream tasks.

Our findings suggest that predictions based on MAEs for fully sampled images represent the \textbf{upper bound} of performance comparison. Remarkably, even with undersampled k-space data, the results are comparable to those from fully sampled images. This underscores the robustness of KMAE and the potential for direct application in k-space-based diagnostics, cardiac assessment, and other CMR applications.

\noindent \textit{\textbf{Outlook.}} The current work only verified the feasibility of the k-space analysis method with single-coil-acquired CMR data. Future work will extend it to multi-coil CMR scans, thereby allowing for the incorporation of more redundant information and further improvement of the estimation accuracy compared to image-domain-based methods.

\section{Conclusion}

In this study, we introduce KMAE model, designed to utilize k-space data for tasks such as disease classification, phenotype regression, and cardiac segmentation. Our findings reveal that Transformer-based architectures effectively process k-space data, achieving comparable classification and regression performance to image-domain models and successfully emulating image-domain segmentation techniques. Moreover, KMAE maintains consistent performance with undersampled k-space data, underscoring its robustness and potential for accelerated MRI applications. This research also highlights the considerable promise of employing k-space data in cardiac MRI and confirms the suitability of Transformer architectures for such applications. Future research should extend to multi-coil CMR scans and explore further downstream tasks to validate and expand these findings in clinical settings.
\\\\
\noindent \textit{\textbf{Acknowledgements and \discintname}} This research has been conducted using the UK Biobank Resource under Application Number 87802. This work is funded by the European Research Council (ERC) project Deep4MI (884622). The authors have no competing interests to declare that are relevant to the content of this article.

%
%
%
%
\bibliographystyle{splncs04}
\bibliography{ref}

\end{document}